\documentclass[12pt]{book}

\usepackage[dvips]{graphicx,color}
\usepackage{makeidx,tsukuba}

\makeauthorindex
\makeindex

\begin{document}

\BookTitle{\itshape The 28th International Cosmic Ray Conference}
\CopyRight{\copyright 2003 by Universal Academy Press, Inc.}
\pagenumbering{arabic}

\chapter{A Study of the Ground Level Event of April 15, 2001 with GRAND }

\author{Christopher D'Andrea and John Poirier\\
{\it
Center for Astrophysics at Notre Dame, University of Notre Dame,
\\Notre Dame, IN 46556, USA} }

\section*{Abstract}
Project {\small GRAND}, a proportional wire chamber array,
is used to examine the increased
counting rate of single muons during the solar Ground Level Event (GLE) of
April 15, 2001. Data are presented and compared to our background rates and
to that of neutron monitor stations.  This GLE is seen with a statistical
significance of 6.1$\sigma$.

\section{Introduction}
On April 15, 2001 the IPS Radio and Space Services website reported a rare
X14  ($14 \times 10^{-4} W/m^2$ at one AU from the sun) solar flare
with a maximum X-ray intensity at 13.83 UT.
The flare originated on the sun at coordinates S20 W85.
During the period from 1976 - 1998, only sixteen flares of
magnitude X10 or greater were reported;
observations on some of these high energy
flares are given in [1,4,6,14].

During the peak time
of the flare, the sun was at an altitude of $35^\circ$
and an azimuth of $110^\circ$ in a local coordinate system.
Even though the sun is near
the minimum detectable elevation angle (27$^\circ$),
the interplanetary magnetic field (IMF) and the magnetic field of the
earth
alter the direction of charged particles from the sun allowing their charged
secondaries  to be more easily
detected by {\small GRAND}.  Muon telescopes are typically sensitive to
higher primary energies than neutron monitors.

Charged particles emitted from the sun spiral around the magnetic field
lines which themselves curve in an Archimedes spiral.  Therefore the GLE
associated with a particular X-ray flare should occur at a slightly
later time than the X-ray event (due to the longer path length and slower
velocity) and at a different angle than the sun, thus providing better
detector
acceptance even though a direct line to the sun is at a low elevation angle
(see
also [8]).

Further details on the observation of this GLE can be found in [12].

\section{Data and Observations}

Project {\small GRAND} is an array of 64 proportional wire stations located
at
$41.7^\circ$~N
 and $86.2^\circ$~W at an altitude of 220~m above sea level.  The array is
used
to detect ground level charged particles produced from primaries undergoing
hadronic interactions.
Each station contains eight 1.29 m$^2$ proportional wire chamber (PWC)
planes.
The position sensitive PWC planes allow the direction
of a muon track to be measured to within 0.26$^\circ$, on average, in each
of
two
projected planes.
A 50~mm thick steel plate is located above the bottom two PWC planes.
Muon tracks can thus be distinguished from electrons or hadrons.
The array collects data at a rate of $\sim$2400 muons per second.
Additional
details are on the webpage: {\ttfamily http://www.nd.edu/$\sim$grand}.

{\small GRAND} recorded a continuous data file on magnetic tape
containing information
from  April 14, 2001 at 8:40~UT through April 16 at 21:30~UT, providing
background
information before and after the flare time.
The counting rate from 14.5 hours to 24.5
hours UT was examined in 0.1 hour bins.
Deviations (r.m.s.) from average counting
rate were measured individually for each station of the array.
In order to prevent
spurious individual station variations from giving a possible signal in
the sum-of-all-stations, one-third of the stations with the highest
individual
r.m.s.\  deviations were eliminated from consideration leaving 39 stations
(an
extremely severe
cut to ensure with great confidence that erratic station behavior could
not cause a false signal).

The six bins of time from 14.0 to 14.6 hours UT were used as a signal time;
the bins from 9.5 to 19.5 hours UT (minus the signal time) were
used as background.
Since it is known that the muon rate depends slightly on the time of day
[11] (due to small
solar effects in air pressure, temperature, and in the IMF),
the counting rate for the
background time period was fit to a quadratic curve.  The muon counting rates
for
April 14 and April 16 (the preceeding and the
next day from the signal) were examined and
it was verified that the backgrounds from these two days are
similar to the curve for April 15;
the average of four years of data also shows a similar shape.
The data as percentage above
background are shown in Figure 1 with error bars.
For comparison, the data from neutron
monitors at Newark [3] and Oulu [9] during this GLE are
also shown.  Our muon data rate has not been corrected for changes in
pressure.
However, the pressure was recorded every half hour
and found to be constant for the entire interval of time analyzed to
within
$\pm$0.15\% and thus could not cause the excess counting rate observed.
The amount of signal above background yields a GLE signal
with a significance of 6.1$\sigma$ (where $\sigma$ is the total error on the
signal and background).
 \begin{figure}[t]
  \begin{center}
    \includegraphics[width=5.8in]{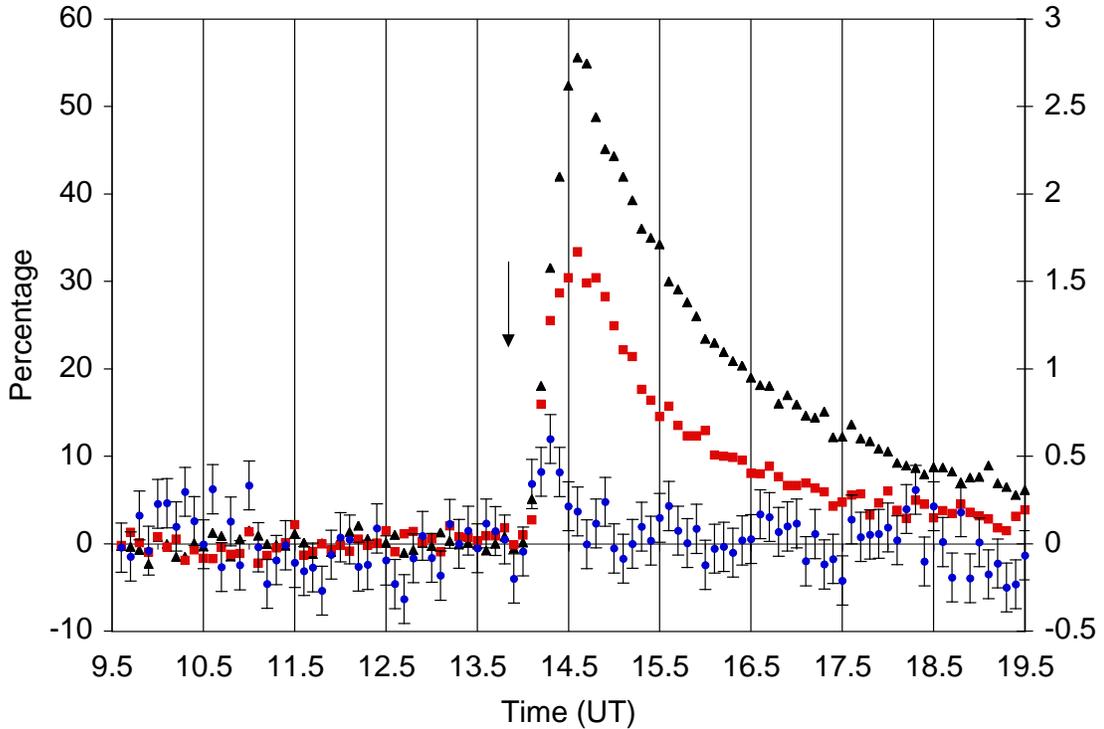}
  \end{center}
  \vspace{-0.5pc}
  \caption{Counting rate increase (in percent) above background.  The left
scale
is used for the neutron monitor stations (squares and triangles); the right
scale is for Project {\small GRAND} (circles with error bars).
The arrow shows the time the X-Ray flare occurred on the sun (0.13 light
hours
away).}
\vspace{-0.5pc}
\end{figure}

Data from the Newark [3], and
Oulu [9] Neutron Monitor Stations are compared in Figure 1, the data for
Climax [5]
are very close to that of Newark and for clarity are not shown.
The energy thresholds for the Climax, Newark,
and Oulu neutron monitor stations are influenced by their vertical
geomagnetic
cutoff rigidity (3.0, 2.1, and 0.8 GV respectively) and overburden of air
(their
corresponding heights above sea level are 3400, 50, and 15~m).
The corresponding numbers for {\small GRAND} are a geomagnetic cutoff
rigidity
of 1.9~GV and a height  of 220~m above sea level.
However, our primary energy is also influenced by the different
mechanisms which produce muons (rather than neutrons) which
subsequently reach detection level.
While looking at single muon tracks at ground level, our threshold primary
proton energy is 3 - 10 GeV [12], depending on the primary spectral index.
Thus our peak is smaller in amplitude due to lower fluxes higher primary
energies.
The Newark signal has a half-height time at 14.21
hours, the Climax signal at 14.16 hours, and Oulu at 14.27 hours
(where the half-height time is
the time at which the data has risen above background to half
of the highest measured data point of the GLE).
A similar calculation for our half-height time is 14.09 hours
which is $\sim$ 0.1 hour earlier
(the earlier time might be expected on the basis of higher primary
energies compared to the neutron monitors [2] with faster protons traveling
at
smaller pitch angles yielding shorter path lengths.
The width (full width at half-of-maximum above background)
of the Climax, Newark, and Oulu peaks are 1.27, 1.24, and 1.61 hours in
contrast to our value of 0.48 hour.
Our peak is thus $\sim$0.9 hours shorter in duration (typically,
higher energies have a shorter duration than lower energies as is seen in
satellite data for protons at lower energies.

\section{Conclusions}
Project {\small GRAND} sees a ground level signal with a significance of
6.1$\sigma$
when
examining the secondary muon counting rate at ground level between 14.0 and
14.6
hours UT on April 15, 2001. This signal is obtained with no restrictions on
the
angle of the muons, consistent with protons originating at the surface of the
sun
and accelerated during the time of the X14 X-ray flare.
Our data at $\pm5^{\circ}$  angles from the direction of the sun show no
increase;
therefore our
detection is not due to gamma rays or neutrons from the sun.
Our GLE detection
occurs $\sim$0.1 hour earlier than Climax, Newark, or Oulu neutron monitor
signals and is shorter in duration
which might be expected considering the higher energies involved.

The authors wish to thank M. Dunford, J. Vermedahl, and M. L\'opez del
Puerto.  Thanks to L. Miroshnichenko for preliminary
intensity calculations and M. Shea and D. Smart for calculations of
our asymptotic direction as a function of rigidity; The GOES science team, J.
Clem and R. Pyle (Bartol Neutron Monitor Program funded
through NSF Grant ATM-0000315), C. Lopate
(Climax Neutron Monitor operated by The University of Chicago and funded
through  NSF grant ATM-9912341), and the Oulu
Neutron Monitor for the use of their data.
{\small GRAND} is funded through the University of Notre Dame and private
grants.

\section{References}
\re
\ 1. Bieber J.\ et al.\ 2002,  ApJ 567, 622
\re
\ 2. Clem J. and Dorman L. \ 2000, Space Science Review 93, 335
\re
\ 3. Clem J. and Pyle R. \ 2002, {\it Private Communication}
\re
\ 4. Falcone A., Ryan J., \ 1999, Astroparticle Physics 11, 283
\re
\ 5. Lopate C. \ 2002, {\it Private Communication}
\re
\ 6. Miroshnichenko L. \ 2001, Solar Cosmic Rays, (Kluwer, Norwell, MA)
\re
\ 7. Miroshnichenko L. \ 2002, {\it Private Communication}
\re
\ 8. Munakata K. \ et al. \ 2001, Proceedings of the 27th International
Cosmic Ray Conferece (Hamburg), 3494
\re
\ 9. Oulu Neutron Monitor Website \ 2002,
{\ttfamily http://spaceweb.oulu.fi/projects/crs}
\re
\ 10. Poirier J., Roesler S., Fasso A.\ 2002, Astroparticle Physics
17, 441
\re
\ 11. Poirier J., D'Andrea C. \ 2001, Proceedings of the 27th
International Cosmic Ray Conferece (Hamburg), 3923 
\re
\ 12. Poirier J., D'Andrea C. \ 2002, Journal of Geophysical
Research 107, 1376
\re
\ 13. Shea M. and Smart D. \ 2002, {\it Private Communication}
\re
\ 14. Swinson D. B., Shea M. A.\ 1990, Geophysical Research Letters
8, 1073
\endofpaper
\end{document}